\documentclass{PoS}

\usepackage[utf8x]{inputenc}

\title{Forward particle production in proton-nucleus collisions at NLO}

\ShortTitle{Forward particle production in proton-nucleus collisions at NLO}

\author{\speaker{Bertrand Ducloué} \\
	Institut de physique théorique, Université Paris Saclay, CEA, CNRS, F-91191 Gif-sur-Yvette, France \\
	E-mail: \email{bertrand.ducloue@ipht.fr}}

\abstract{Reaching next-to-leading order accuracy in perturbative calculations of particle production in QCD at high energy is essential for reliable phenomenological applications. In recent years, the Color Glass Condensate effective theory (the natural framework for such calculations) has indeed been promoted to NLO accuracy. However, the first NLO calculation of single-inclusive hadron production met with an unexpected difficulty: the cross-section suddenly turns negative at transverse momenta of the order of a few GeV, in a range where perturbation theory is expected to be reliable. We summarize recent efforts to understand and solve this issue, as well as to develop a running coupling scheme that can be used to consistently describe various processes in this formalism.}

\FullConference{XIII Quark Confinement and the Hadron Spectrum - Confinement2018\\
		31 July - 6 August 2018\\
		Maynooth University, Ireland}

\usepackage{graphicx}
\usepackage{bm,amsmath,amssymb}
\usepackage[mathscr]{eucal}
\usepackage{color}
\usepackage{afterpage}

\newcommand{\nn}{\nonumber\\}
\newcommand{\dif}{{\rm d}}
\newcommand{\rmd}{{\rm d}}
\newcommand{\rme}{{\rm e}}
\newcommand{\rmi}{{\rm i}}
\newcommand{\rmP}{{\rm P}}

\newcommand{\rmtr}{{\rm tr}}

\newcommand{\del}{\partial}

\newcommand{\bk}{\bm{k}}
\newcommand{\bq}{\bm{q}}

\newcommand{\bel}{\bm{\ell}}
\newcommand{\bx}{\bm{x}}
\newcommand{\by}{\bm{y}}

\newcommand{\bz}{\bm{z}}

\newcommand{\br}{\bm{r}}

\newcommand{\abar}{\bar{\alpha}_s}

\newcommand{\ssE}{{\rm \scriptscriptstyle E}}

\begin{document}

\section{Introduction}

Forward particle production at semi-hard transverse momenta in high energy proton-proton and proton-nucleus collisions can provide valuable information about the nuclear wavefunction at small $x$ in a domain where the strong coupling is small enough for a perturbative treatment to be applicable but where gluon occupation numbers are high and non-linear effects such as saturation are expected to be important. To extract as much information as possible from measurements performed at RHIC and the LHC, accurate theoretical predictions are important. In the recent years several works have been devoted to push the Color Glass Condensate (CGC) effective theory~\cite{Gelis:2010nm}, which is the natural framework to deal with such processes within perturbative QCD, to NLO accuracy. This includes the computation of NLO corrections to the non-linear equations governing the evolution of scattering amplitudes at high energy (the B-JIMWLK hierarchy of equations~\cite{Balitsky:1995ub,JalilianMarian:1997jx,JalilianMarian:1997gr,Kovner:2000pt,Iancu:2000hn,Iancu:2001ad,Ferreiro:2001qy} and its mean field approximation, the Balitsky-Kovchegov equation~\cite{Balitsky:1995ub,Kovchegov:1999yj}) as well as to the impact factors describing the coupling between the dilute projectile and the dense target's gluon distribution. However the first studies implementing these corrections found that they can lead to unphysical results.
In the case of the BK equation, it was found that the NLO corrections~\cite{Balitsky:2008zza,Balitsky:2013fea,Kovner:2013ona} make the evolution unstable~\cite{Avsar:2011ds,Lappi:2015fma} because of large collinear logarithms which need to be resummed to all orders~\cite{Beuf:2014uia,Iancu:2015vea,Iancu:2015joa}. This resummation indeed leads to stable and physical results~\cite{Lappi:2016fmu}.
Regarding impact factors, the first NLO calculations of forward particle production in proton-nucleus collisions~\cite{Chirilli:2011km,Chirilli:2012jd,Stasto:2013cha,Stasto:2014sea,Xiao:2014uba,Watanabe:2015tja,Ducloue:2016shw} obtained cross sections which suddenly turn negative when the transverse momentum of the produced hadron becomes of the order of a few GeV. This is a particularly surprising result since the formalism is expected to apply in this semi-hard region. Here we will summarize the recent efforts to understand and solve this issue~\cite{Ducloue:2016shw,Ducloue:2017mpb,Ducloue:2017dit}.

\section{Forward particle production at leading and next-to-leading order}

Let us first summarize the basic formulas that will be useful in the following sections. Since we consider particle production at forward rapidities, we work in the so-called ``hybrid factorization''~\cite{Dumitru:2005gt} where the projectile proton, being probed at rather large $x$ values, can be described by usual collinear parton distribution functions (PDFs). The target nucleus, on the other hand, is probed at very small $x$ values and is described in terms of ``unintegrated'' gluon distributions (UGDs) encoding the information about gluon densities as a function of $x$ and transverse momentum. The non-linear evolution of the UGD as a function of $x$ is governed by the Balitsky-Kovchegov (BK) equation~\cite{Balitsky:1995ub,Kovchegov:1999yj}.

\subsection{Leading order}

In the hybrid factorization, the LO quark multiplicity reads
\begin{align}
\label{sigmaLO}
\frac{\dif N^{{\rm LO}}}{\dif^2\bk\, \dif \eta} 
=\, \frac{x_p q(x_p)}{(2\pi)^2}\,
\mathcal{S}(\bk,X_g)\,,
\end{align}
where $\bk$ is the transverse momentum of the produced quark, $\eta$ its rapidity, $q$ is the collinear quark distribution in the projectile proton and $\mathcal{S}$ is the ``unintegrated gluon distribution'' in the target nucleus. Note that for simplicity, throughout this work we will only consider the quark channel $q \to q$ as it exhibits the same issues as the total multiplicity. In addition we do not consider the fragmentation of quarks into hadrons, which would be necessary for phenomenological studies but is not needed here. The longitudinal momentum fractions $x_p$ and $X_g$ probed respectively in the proton and nucleus are
\begin{equation}
\label{Xg}
x_p=\frac{k_\perp}{\sqrt{s}}\,\rme^\eta\,,\qquad X_g=\frac{k_\perp}{\sqrt{s}}\,\rme^{-\eta}=\frac{k_{\perp}^2}{x_p s}\,,
\end{equation}
where $k_{\perp}=|\bk|$ and $\sqrt{s}$ is the center of mass energy. From these expressions we see that at high energy and forward rapidity we are in the regime $X_g\ll x_p < 1$. The unintegrated gluon distribution $\mathcal{S}$ in~(\ref{sigmaLO}) is defined as the Fourier transform of the $S$-matrix  describing the elastic scattering between a small color dipole and the nucleus:
\begin{equation}
\mathcal{S}(\bk,X) = \int \dif^2\br\, 
\rme^{-\rmi \bk \cdot \br}
S(\br, X),
\end{equation}
with
\begin{equation}
\label{scoord}
S(\bx,\by;X) = \frac{1}{N_c}
\left\langle \rmtr \left[V(\bx) V^{\dagger}(\by) \right]\right\rangle_{X}\,,
\end{equation}
where $V$ is Wilson line in the color field of the target in the fundamental (since we are considering the scattering of a quark) representation:
\begin{equation}
\label{wline}
V(\bx) = \rmP \left[\rmi g \int \dif x^+ A^-_a(x^+,\bx) t^a  \right].
\end{equation}
Here $g$ is the strong coupling, $x^+$ the light-cone time of the quark and $\bx$ its transverse coordinate, $A^-_a$ is the relevant component of the color field of the target nucleus and $t^a$ are the generators of SU($N_c$) in the fundamental representation.

Since we work in the small $X_g$ limit, we need to resum to all orders contributions enhanced by powers of $\abar Y_g$, with $\abar\equiv \alpha_s N_c/\pi$ and $Y_g \equiv \ln(1/X_g)$. These contributions are generated by the high energy evolution and correspond to the successive emission of soft gluons with strongly ordered longitudinal momenta. To leading order in pQCD and in the mean field approximation suitable for a dense target and a large number of colors, the high energy evolution of $S$ is governed by the Balitsky-Kovchegov (BK) equation~\cite{Balitsky:1995ub,Kovchegov:1999yj}:
\begin{align}\label{BK}
X \frac{\del }{\del X} \,S(\bx,\by; X)=
\frac{\abar}{2\pi}\, \int \rmd^2\bz\,
\frac{(\bx-\by)^2}{(\bx-\bz)^2(\bz-\by)^2}\,\Big[
S(\bx,\bz; X) S(\bz, \by; X)
-S(\bx,\by; X)\Big]\,,
\end{align}
where $(\bx,\by)$ are the coordinates of the parent dipole which scatters off the target and $\bz$ is the transverse coordinate of the soft gluon emitted in one step of the evolution. Starting from an initial condition, for example the McLerran-Venugopalan (MV) model~\cite{McLerran:1993ni,McLerran:1993ka}, formulated at a given $X_0$, it is then possible to evolve $S$ to lower values of $X$.

\subsection{Next-to-leading order}
\label{sec:NLO}

When going to next-to-leading order, one should include the contributions suppressed by an additional power of $\alpha_s$ which is not enhanced by an $Y_g$ factor. As mentioned in the introduction, there are two sources of such contributions: the NLO corrections to the high energy BK evolution, which are process-independent, and the ones to the impact factor, or hard part, of the process. The impact factor describes the scattering in the absence of any high energy evolution, i.e. at leading order it corresponds to the scattering of a bare quark off an unevolved target with $X=X_0$. At next-to-leading order one has to include contributions where the quark emits a gluon, and thus to consider the scattering of a quark-gluon system off the nucleus. However, only the case where this additional gluon is relatively hard should be considered to be part of the NLO corrections. Indeed, these are the contributions not enhanced by a power of $Y_g$ and thus not already included in the LO expression~(\ref{sigmaLO}) via the high energy evolution. Instead of doing this explicit separation between leading and next-to-leading order contributions, as was done in~\cite{Chirilli:2011km,Chirilli:2012jd} where the NLO impact factor for single inclusive particle production was computed, we will rather first consider an ``unsubtracted'' expression for the NLO quark multiplicity~\cite{Iancu:2016vyg} which mixes both types of contributions. The reason for this and the relation with the formulas obtained in~\cite{Chirilli:2011km,Chirilli:2012jd} will be discussed in Sec.~\ref{sec:subunsub}. Following~\cite{Iancu:2016vyg}, the NLO quark multiplicity can be written as
\begin{align}
\label{nlonc}
\frac{\dif N^{{\rm NLO}}}{\dif^2\bk\, \dif \eta} 
=\, &
\frac{x_p q(x_p)}{(2\pi)^2}
\mathcal{S}(\bk,X_0)	+
\frac{1}{4\pi}\int_0^{1-X_g/X_0} \dif \xi\,
\frac{1+\xi^2}{1-\xi}
\nn
& \times \bigg[\Theta(\xi-x_p)\frac{x_p}{\xi}\,
q\left(\frac{x_p}{\xi}\right) \left(\frac{2 C_{\rm F}}{N_c} \mathcal{I}(\bk,\xi,X(\xi)) +
\mathcal{J}(\bk,\xi,X(\xi)) \right) \nn
& \hspace{6mm} - x_p q(x_p) \left(\frac{2 C_{\rm F}}{N_c} \mathcal{I}_v(\bk,\xi,X(\xi))+ \mathcal{J}_v(\bk,\xi,X(\xi)) \right) \bigg],
\end{align}
with the integrals
\begin{align}
\label{I}
\mathcal{I}(\bk,\xi,X(\xi)) &=
\abar \int \frac{\dif^2\bq}{(2\pi)^2}
\left[\frac{\bk-\bq}{(\bk-\bq)^2}
- \frac{\bk-\xi \bq}{(\bk-\xi \bq)^2} \right]^2 
\mathcal{S}(\bq, X(\xi)),
\\
\label{Iv}
\mathcal{I}_v(\bk,\xi,X(\xi)) &=
\abar \int \frac{\dif^2\bq}{(2\pi)^2}
\left[\frac{\bk-\bq}{(\bk-\bq)^2}
- \frac{\xi \bk-\bq}{(\xi\bk-\bq)^2} \right]^2 
\mathcal{S}(\bk, X(\xi)), \\
\label{J}
\mathcal{J}(\bk,\xi,X(\xi)) &=
\abar \int \frac{\dif^2\bq}{(2\pi)^2}
\mathcal{S}(\bq, X(\xi))\bigg[
\frac{2(\bk-\xi \bq)\!\cdot\!(\bk-\bq)}{(\bk-\xi \bq)^2(\bk-\bq)^2} \nn
& \hspace{44mm} - \int \frac{\dif^2\bel}{(2\pi)^2}
\frac{2(\bk-\xi \bq)\!\cdot\!(\bk-\bel)}{(\bk-\xi \bq)^2(\bk-\bel)^2}
\mathcal{S}(\bel, X(\xi))\bigg],
\\
\label{Jv}
\mathcal{J}_v(\bk,\xi,X(\xi)) &=
\abar \int \frac{\dif^2\bq}{(2\pi)^2}
\mathcal{S}(\bk, X(\xi))\bigg[
\frac{2(\xi\bk-\bq)\!\cdot\!(\bk-\bq)}{(\xi\bk- \bq)^2(\bk-\bq)^2} \nn
& \hspace{44mm} - \int \frac{\dif^2\bel}{(2\pi)^2}
\frac{2(\xi\bk-\bq)\!\cdot\!(\bel-\bq)}{(\xi\bk-\bq)^2(\bel-\bq)^2}
\mathcal{S}(\bel, X(\xi))\bigg].
\end{align}
In these expressions, $1-\xi$ is the longitudinal momentum fraction of the incoming quark taken by the primary gluon, i.e. the limit $\xi \to 1$ corresponds to a soft gluon. In this ``unsubtracted'' formulation, there is no explicit separation between the LO and NLO contributions: the first term in the right hand side of~(\ref{nlonc}) corresponds to the tree-level result without high energy evolution, thus $\mathcal{S}$ is evaluated at the initial condition $X_0$. The second term contains both the high energy evolution and the pure $\alpha_s$ corrections to the impact factor. In particular, the second line in~(\ref{nlonc}) contains the real contributions, which are thus weighted by the quark distribution evaluated at $x_p/\xi$, while the third line contains the virtual ones, weighted by $q(x_p)$. Both real and virtual contributions can be written as the sum of a term proportional to the $C_{\rm F}$ color factor (namely $\mathcal{I}$ or $\mathcal{I}_v$) and a term proportional to $N_c$ ($\mathcal{J}$ or $\mathcal{J}_v$). The $S$-matrices in these terms are evaluated at the rapidity scale $X(\xi)$. This scale depends on $\xi$ since the emission of a primary gluon with a rapidity $y=\ln(1/(1-\xi))$ only leaves an interval $Y_g-y= \ln((1-\xi)/X_g)$ for the high energy evolution. Therefore we write
\begin{equation}
X(\xi) = \frac{X_g}{1-\xi} = \frac{k_{\perp}^2}{x_p s (1-\xi)}\, .
\end{equation}
This expression is correct as long as $k_{\perp} \gtrsim Q_s(X)$, where $Q_s(X)$ is the target saturation scale (see the discussion in~\cite{Iancu:2016vyg}). This is indeed the kinematics we are interested in here, since it corresponds to the region where the results obtained in~\cite{Stasto:2013cha} become unphysical. Note that the upper limit on the $\xi$ integral in~(\ref{nlonc}) enforces the restriction $X(\xi) \le X_0$, with $X_0$ the initial condition for the BK evolution.

In~(\ref{nlonc}), the high energy evolution is contained in the terms proportional to $N_c$ (which also contain a part of the NLO corrections). Indeed, it is easy to check that the terms proportional to $C_{\rm F}$ vanish in the soft gluon limit $\xi\to 1$, therefore the integral over $\xi$ in these terms do not produce a large logarithm of $X_0/X_g$. The $C_{\rm F}$ terms, however, contain collinear divergences (at $\bq=\bk$ and $\bq=\bk/\xi$ in the real term and at $\bq=\bk$ and $\bq=\xi\bk$ in the virtual term) which can be absorbed into the DGLAP evolution of quark distributions and fragmentation functions. This is done in~\cite{Chirilli:2011km,Chirilli:2012jd} using dimensional regularization, and one should then replace $\mathcal{I}$ and $\mathcal{I}_v$ with the respective finite expressions:
\begin{align}
\label{Ifin}
\mathcal{I}^{\rm fin}(\bk,\xi,X(\xi)) & =
\abar\! \int\! \frac{\dif^2\br}{4\pi}\,
S(\br, X(\xi))\ln \frac{c_0^2}{\br^2 \mu^2}
\left(\rme^{-\rmi \bk \cdot \br} 
+ \frac{1}{\xi^2} \rme^{-\rmi {\textstyle\frac{\bk}{\xi}} \cdot \br} 
\right) \hspace{4cm} \nn
& \hspace{4mm} - 2 \abar \!\int\! \frac{\dif^2 \bq}{(2\pi)^2}
\frac{(\bk - \xi \bq)\cdot(\bk-\bq)}{(\bk - \xi \bq)^2(\bk-\bq)^2}
\mathcal{S}(\bq, X(\xi)), \\
\label{Ivfin}
\mathcal{I}_v^{\rm fin}(\bk,\xi,X(\xi)) & =
\abar
\left[\ln\frac{\bk^2}{\mu^2} + \ln(1-\xi)^2 \right]
\frac{\mathcal{S}(\bk, X(\xi))}{2\pi},
\end{align}
where $\mu$ is the factorization scale and $c_0 = 2 \rme^{-\gamma_\ssE}$.

\section{Unsubtracted, subtracted and CXY expressions}
\label{sec:subunsub}

As written in~(\ref{nlonc}), the expression for the NLO multiplicity is not an explicit sum of LO and NLO contributions. Furthermore, it is non-local in rapidity since the $S$-matrices in the last two lines of this expression are evaluated at the floating scale $X(\xi)$. This is in contrast to the original expression by Chirilli, Xiao, and Yuan (CXY)~\cite{Chirilli:2011km,Chirilli:2012jd}, in which the LO and NLO contributions were separated and where the $S$-matrices were all evaluated at the LO value $X_g$. We will now detail the manipulations and approximations that relate these two formulations.

As explained in the previous section, the ``$N_c$-terms'' involving $\mathcal{J}$ and $\mathcal{J}_v$ in~(\ref{nlonc}) contain both the LL high energy evolution and the fixed order NLO corrections proportional to $N_c$. Therefore we will first consider the sum of the leading order term and the $N_c$-terms from the NLO contributions. On the other hand, the ``$C_{\rm F}$-terms'' involving $\mathcal{I}$ and $\mathcal{I}_v$ in~(\ref{nlonc}) are not related to the high energy evolution and are thus pure NLO corrections that will be added at the end.

\begin{figure}[t]
\includegraphics[width=0.48\textwidth]{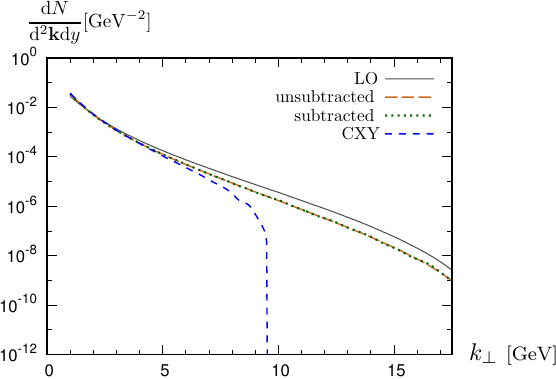}
\hspace{3mm}
\includegraphics[width=0.48\textwidth]{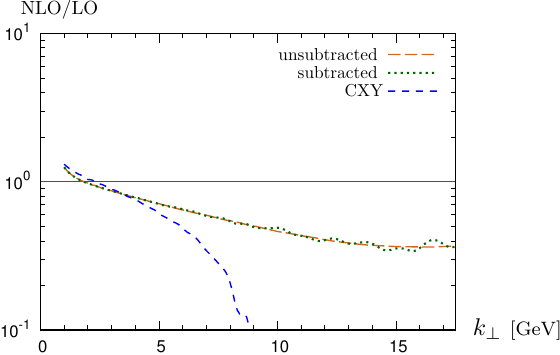}
\caption{NLO quark channel multiplicity at fixed coupling for $\sqrt{s}=500$ GeV and a rapidity $y=3.2$, using the unsubtracted~(\ref{nlounsub}), subtracted~(\ref{nlosub}) and CXY~(\ref{nlocxy}) formulations. On the left the multiplicity and on the right the NLO/LO ratio. Figure from~\cite{Ducloue:2017mpb}.}
\label{fig:fcBK_NLO}
\end{figure}

Let us first introduce simplified notations. We can rewrite the sum of LO and $N_c$ NLO contributions in~(\ref{nlonc}) as
\begin{equation}
\label{nlounsub}
\frac{\dif N^{{\rm LO} + N_c}}{\dif^2\bk\, \dif \eta}  = 
\frac{x_p q(x_p)}{(2\pi)^2}\,
\mathcal{S}(\bk,X_0)
+\int_0^{1-X_g/X_0} 
\frac{\dif \xi}{1-\xi} \, \mathcal{K}(\bk,\xi,X(\xi))
\equiv \frac{\dif N^{\rm IC}}{\dif^2\bk\, \dif \eta}
+\frac{\dif N^{N_c,{\rm Unsub}}}{\dif^2\bk\, \dif \eta},
\end{equation}
where we have defined
\begin{equation}
\mathcal{K}=\frac{1}{4\pi}(1+\xi^2)\left[\Theta(\xi-x_p)\frac{x_p}{\xi}\,
q\left(\frac{x_p}{\xi}\right)\mathcal{J}(\bk,\xi,X(\xi))- x_p q(x_p) \mathcal{J}_v(\bk,\xi,X(\xi))\right].
\end{equation}
The first term in the right hand side of~(\ref{nlounsub}) is the tree-level contribution without high energy evolution (i.e. the initial condition, or IC), and the ``Unsub'' superscript in the second term stands for ``unsubtracted''. As explained previously, this last term contains the emission of a first primary gluon with arbitrary kinematics, plus any number of further emissions treated in the eikonal approximation. Therefore, in the limit where the primary gluon is soft (which corresponds to $\xi \to 1$) we should recover the LO result~(\ref{sigmaLO}). Indeed, this can be checked easily by using the integral representation of the BK equation in momentum space:
\begin{equation}
\label{bkmom}
\mathcal{S}(\bk,X_g) = 
\mathcal{S}(\bk,X_0)
+2 \pi \int_{X_g}^{X_0}
\frac{\dif X}{X}\, 
\big[
\mathcal{J}(\bk,\xi=1,X) - \mathcal{J}_{v}(\bk,\xi=1,X) 
\big].
\end{equation}
This allows us to rewrite~(\ref{nlounsub}) in the following way:
\begin{align}
\label{nlosub}
\frac{\dif N^{{\rm LO} + N_c}}{\dif^2\bk\, \dif \eta} & = 
\frac{x_p q(x_p)}{(2\pi)^2}\,
\mathcal{S}(\bk,X_g)
+\int_0^{1-X_g/X_0} 
\frac{\dif \xi}{1-\xi} \, \big[\mathcal{K}(\bk,\xi,X(\xi))
-\mathcal{K}(\bk,\xi=1,X(\xi))\big] \nn
& \equiv \frac{\dif N^{\rm LO}}{\dif^2\bk\, \dif \eta}
+\frac{\dif N^{N_c,{\rm Sub}}}{\dif^2\bk\, \dif \eta}.
\end{align}
This last expression, where the ``Sub'' superscript stands for ``subtracted'', is an explicit sum of the LO contribution~(\ref{sigmaLO}) and a fixed order NLO correction which develops no small$-X_g$ logarithm since the integrand is vanishing in the limit $\xi \to 1$. Eq.~(\ref{nlosub}) might seem more natural than~(\ref{nlounsub}) from a perturbative expansion point of view, however because it involves adding and subtracting a large contribution one can expect it to be less stable in numerical evaluations. This is confirmed in Fig.~\ref{fig:fcBK_NLO}, where the multiplicity and NLO/LO ratio are shown as a function of transverse momentum using either the ``unsubtracted'' or ``subtracted'' formulations: at large transverse momentum the later one shows small oscillations due to numerical inaccuracies.

We see in Fig.~\ref{fig:fcBK_NLO} that the NLO multiplicity using either~(\ref{nlounsub}) or~(\ref{nlosub}) is positive even at large transverse momentum, contrary to the results obtained in~\cite{Stasto:2013cha} using the ``CXY'' expressions presented in~\cite{Chirilli:2011km,Chirilli:2012jd}. To arrive at the ``CXY'' expressions, we need to make some approximations in~(\ref{nlosub}). The first one is to replace the rapidity argument of the $S$-matrices in the second term by the LO value $X_g$. The second one is to neglect $X_g/X_0 \ll 1$ in the upper limit of the $\xi$ integral and therefore to replace this limit by 1. We thus arrive at
\begin{align}
\label{nlocxy}
\frac{\dif N^{{\rm LO} + N_c}}{\dif^2\bk\, \dif \eta}\bigg|_{\rm CXY} & = 
\frac{x_p q(x_p)}{(2\pi)^2}\,
\mathcal{S}(\bk,X_g)
+ \int_{0}^1
\frac{\dif \xi}{1-\xi} \, \big[\mathcal{K}(\bk,\xi,X_g)
-\mathcal{K}(\bk,\xi=1,X_g)\big] \nn
& \equiv \frac{\dif N^{\rm LO}}{\dif^2\bk\, \dif \eta}
+\frac{\dif N^{N_c,{\rm Sub}}}{\dif^2\bk\, \dif \eta}
\bigg|_{\rm CXY},
\end{align}
which is now equivalent to the formulas presented in~\cite{Chirilli:2011km,Chirilli:2012jd}. These two approximations, which make~(\ref{nlocxy}) no longer equivalent to~(\ref{nlounsub}) and~(\ref{nlosub}), are in principle justified at NLO accuracy since the second term in~(\ref{nlosub}) is a pure $\alpha_s$ correction. However we recall that the equivalence between~(\ref{nlounsub}) and~(\ref{nlosub}) relied on a fine cancellation between large contributions. By using the CXY approximation this cancellation is no longer exact and the negative contribution proportional to $\mathcal{K}(\bk,\xi=1,X_g)$ becomes too large in magnitude and overcompensates for the LO piece in $\mathcal{S}(\bk,X_g)$, leading to negative results at large $k_\perp$, as shown in Fig.~\ref{fig:fcBK_NLO}. We thus conclude that to obtain physical results it is necessary to keep the non-local formulation.

\section{Running coupling}

\subsection{Mixed and coordinate space formulations}

\begin{figure}[t]
\includegraphics[width=0.48\textwidth]{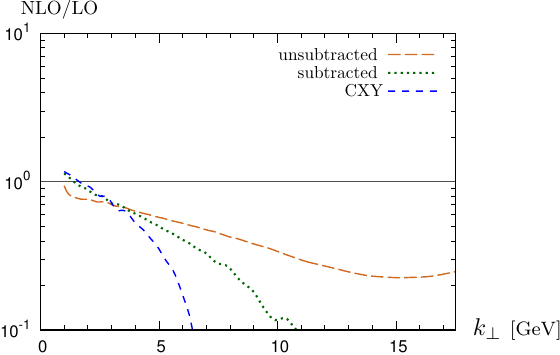}
\hspace{3mm}
\includegraphics[width=0.48\textwidth]{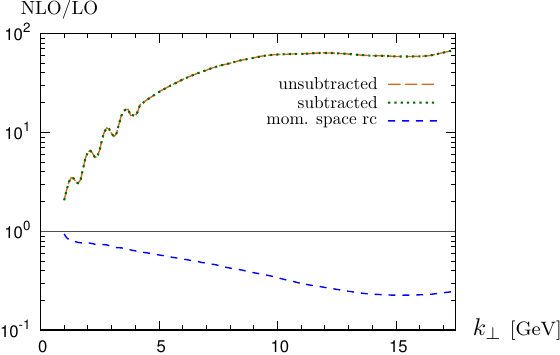}
\caption{NLO/LO ratio for the quark channel multiplicity with running coupling for $\sqrt{s}=500$ GeV and a rapidity $y=3.2$, using the unsubtracted~(\ref{nlounsub}), subtracted~(\ref{nlosub}) and CXY~(\ref{nlocxy}) formulations. Left: Mixed representation with a coordinate space running coupling used when solving the BK equation and a momentum space running coupling for the impact factor. Right: Results obtained when using a coordinate space running coupling both when solving the BK equation and in the impact factor. Figure from~\cite{Ducloue:2017mpb}.}
\label{fig:rcBK_NLO_mom}
\end{figure}

In all the expressions written in the previous sections we have implicitly considered the coupling $\alpha_s$ to be fixed. However the running of the coupling is an important effect which has to be taken into account for realistic phenomenological studies. This leads to an additional complication in the present case: the BK equation is most conveniently written and solved in coordinate space while the multiplicity is written in momentum space. In particular, some coordinate space prescriptions commonly used when solving the BK equation are the smallest dipole and Balitsky's~\cite{Balitsky:2006wa} prescriptions. On the other hand, for the explicit $\alpha_s$ factors appearing in the multiplicity, the most natural choice at semi-hard transverse momenta $k_\perp \gtrsim Q_s(X_g)$ is simply $\abar(k_{\perp})$. If we mix these different couplings in the calculation, some equalities written in the previous sections are no longer true because they relied on the momentum representation~(\ref{bkmom}) of the BK equation. In particular,~(\ref{nlounsub}) will no longer reduce to the correct LO result~(\ref{sigmaLO}) in the eikonal limit $\xi \to 1$, with a mismatch reaching up to 30\%~\cite{Ducloue:2017mpb}. Another consequence is that the ``unsubtracted''~(\ref{nlounsub}) and ``subtracted''~(\ref{nlosub}) expressions are not equivalent anymore, with the later one becoming negative at large transverse momentum due to an oversubtraction (see Fig.~\ref{fig:rcBK_NLO_mom}~(L)). Therefore we are left with an ambiguity: either use~(\ref{nlounsub}) which leads to physical results but does not have the proper LO limit, or use~(\ref{nlosub}) which reduces to the correct LO result but becomes negative at large $k_\perp$.

An attempt to avoid this ambiguity was made in~\cite{Ducloue:2017mpb} by working fully in coordinate space, and performing the Fourier transform of the multiplicity in momentum space at the very end of the calculation. Such an approach allows to use the same coordinate space running coupling everywhere in the calculation. For this we can write $\mathcal{J}$ and $\mathcal{J}_v$ as the following Fourier transforms:
\begin{align}
\label{Jco}
& \mathcal{J}(\bk,\xi,X(\xi)) =  
\int\dif^2 \br\, \rme^{-\rmi \bk \cdot \br}
J(\br,\xi,X(\xi))
\nn
& \hspace{8mm} \equiv  \int \dif^2 \br \,
\rme^{-\rmi \bk \cdot \br}
\int \frac{\dif^2 \bx}{(2\pi)^2}
\,\abar\,
\frac{2 \bx \!\cdot\! (\bx + \br)}{\bx^2(\bx + \br)^2}
\left[ 
S(\br + (1-\xi) \bx,X(\xi))
-S(-\xi\bx,X(\xi)) S(\br + \bx,X(\xi))
\right], \\
\label{Jcov}
& \mathcal{J}_v(\bk,\xi,X(\xi)) =
\int\dif^2 \br\, \rme^{-\rmi \bk \cdot \br}
J_v(\br,\xi,X(\xi))
\nn
& \hspace{24mm} \equiv \!\! \int \!\! \dif^2 \br \,
\rme^{-\rmi \bk \cdot \br} \!\!
\int \!\! \frac{\dif^2 \bx}{(2\pi)^2}
\,\abar\,
\frac{2}{\bx^2}
\left[ 
S(\br - (1-\xi) \bx,X(\xi))
-S(-\bx,X(\xi)) S(\br + \xi \bx,X(\xi))
\right],
\end{align}
and the BK equation in these notations simply reads
\begin{equation}
\label{bkco}
S(\br,X_g)
= S(\br,X_0)
+2 \pi \int\limits_{X_g}^{X_0}
\frac{\dif X}{X}\, 
\left[
J(\br,\xi=1,X) - J_{v}(\br,\xi=1,X) 
\right].
\end{equation}
Using these expressions it becomes possible to use the same coordinate space running coupling both in the impact factor and when solving the BK equation, thus restoring the correct LO limit for the unsubtracted formulation and its equivalence with the subtracted one. The calculation of one-loop running coupling corrections to the BK equation~\cite{Kovchegov:2006wf,Kovchegov:2006vj,Balitsky:2006wa,Balitsky:2008zza} show that they can be minimized by choosing the scale in a way that it reduces to $\abar(r_{\rm min})$, with $r_{\rm min} \equiv {\rm min}\{|\br|, |\bx|,|\br-\bx|\}$, when there is a strong disparity between these three dipoles. This is satisfied in particular by the smallest dipole prescription $\abar(r_{\rm min})$ and by the Balitsky prescription~\cite{Balitsky:2006wa} which is often used in phenomenological studies. In~\cite{Ducloue:2017mpb} the previous expressions were evaluated with a generalization of the Balitsky prescription to $\xi \ne 1$ which led to very troublesome results: while the unsubtracted and subtracted formulations are indeed equivalent, the NLO corrections change sign and the NLO multiplicity can be up to two orders of magnitude larger than the LO one at large $k_\perp$. This is shown in Fig.~\ref{fig:rcBK_NLO_mom}~(R).

\subsection{The fake potential problem and the daughter dipole prescription}

To understand the origin of the issue observed in Fig.~\ref{fig:rcBK_NLO_mom}~(R), we will first consider a simpler example, namely we will compare the following two quantities:
\begin{align}
\label{nk}
&\mathcal{N}_k \equiv \abar(k_{\perp})\, 
\mathcal{S}(\bk)
=\abar(k_{\perp}) 
\int \dif^2 \br\, \rme^{-\rmi \bk \cdot \br} S(\br), 
\\
\label{nr}
&\mathcal{N}_r \equiv 
\int \dif^2 \br\, 
\abar(r_{\perp})
\rme^{-\rmi \bk \cdot \br} S(\br).
\end{align}
Because $\br$ and $\bk$ are Fourier conjugate, one could naively expect that these two quantities would not differ a lot. However if we evaluate these expressions at large $k_\perp$ in the simple McLerran-Venugopalan (MV) model~\cite{McLerran:1993ni,McLerran:1993ka}, we obtain~\cite{Ducloue:2017dit}
\begin{align}
\label{nkres}
\mathcal{N}_k & \simeq \frac{4 \pi \abar(k_{\perp})Q_s^2}{k_{\perp}^4}, \\
\label{nrres}
\mathcal{N}_r & \simeq - \frac{4 \pi}{\bar{b} [\ln (k_{\perp}^2/\Lambda^2)]^2}\,\frac{1}{k_{\perp}^2},
\end{align}
which means that they have opposite signs and different power tails. As shown in~\cite{Ducloue:2017dit}, the result~(\ref{nrres}) is physically incorrect since the $1/k_{\perp}^2$ tail appears because of the singular behavior of $\abar(r_{\perp})$ when $r_\perp \to 0$. Let us stress that this issue, which is due to the fact that the choice of the scale and the Fourier transform do not commute, is related to the UV behavior of the running coupling, i.e. to asymptotic freedom, and not to the way it is regularized in the infrared.

This lack of commutation between the Fourier transform and the running coupling is also the origin of the unphysical results at large $k_\perp$ in Fig.~\ref{fig:rcBK_NLO_mom}~(R). A large $k_\perp$ (compared to $Q_s$) cannot be provided by multiple scattering on the target, therefore it must be balanced by the unobserved gluon. In coordinate space, this means that the dominant contribution to the multiplicity must come from the region $x_{\perp} \sim r_{\perp}$. This physical condition is satisfied if one considers a fixed coupling or a momentum space running coupling
$\alpha_s(k_\perp)$, but not also for a running coupling, like $\alpha_s(r_\perp)$, which depends on $r_\perp$. To see this, consider the contribution from the complementary region at $x_{\perp} \gg r_{\perp}$, which can be estimated as~\cite{Ducloue:2017dit}
\begin{equation}
\label{jxggr}
\mathcal{J}(\bk,\xi) \sim
\int \dif^2 \br \,\frac{\abar}{2\pi^2}  \,
\rme^{-\rmi \bk \cdot \br}
\int_{r_\perp} \frac{\dif^2 \bx }{\bx^2}\,
\left[ S((1-\xi) \bx) - S(-\xi\bx) S(\bx)
\right] \quad \mbox{\rm for} \;\; x_{\perp} \gg r_{\perp} \,.
\end{equation}
The combination of dipole $S$-matrices within the square brackets grows like $x_\perp^2$ for small $x_\perp \sim r_\perp$, while it exponentially vanishes for larger  $x_\perp \gtrsim 1/Q_s$. Accordingly the integral over $\bx$ is dominated by large values  $x_\perp \sim 1/Q_s$ and thus it is independent of its lower limit $r_{\perp}$ in the approximation of interest. So long as the coupling $\abar\equiv\alpha_s N_c/\pi$ is independent of $r_\perp$, the final Fourier transform yields a vanishing result and thus the dominant contribution to the integral comes from the region $x_{\perp} \sim r_{\perp}$ as expected on physical grounds. On the contrary, if the coupling is chosen to depend on $r_{\perp}$, this dependence will lead to a large and unphysical contribution from the region $x_{\perp} \gg r_{\perp}$, i.e. from soft primary gluons.

Following this discussion one could wonder why similar issues do not appear when solving the BK equation with a coordinate space running coupling since the BK equation can be written using the same integrals~(\ref{bkco}). The crucial difference here is that~(\ref{bkco}) involves the difference between $J$ and $J_v$, and in this combination the spurious contributions coming from the region $x_{\perp} \gg r_{\perp}$ cancel as shown in~\cite{Ducloue:2017dit}. In the case of the NLO multiplicity the real and virtual terms are weighted by different PDFs as shown in~(\ref{nlonc}) which prevents this cancellation.

\begin{figure}[t]
\includegraphics[width=0.48\textwidth]{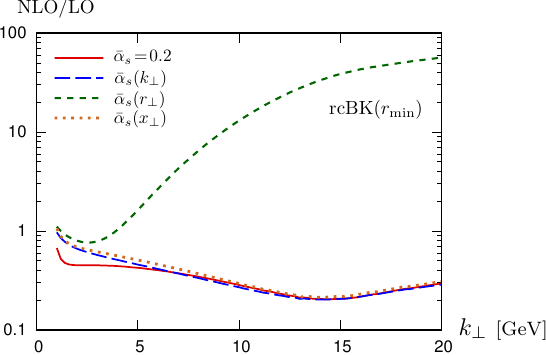}
\hspace{3mm}
\includegraphics[width=0.48\textwidth]{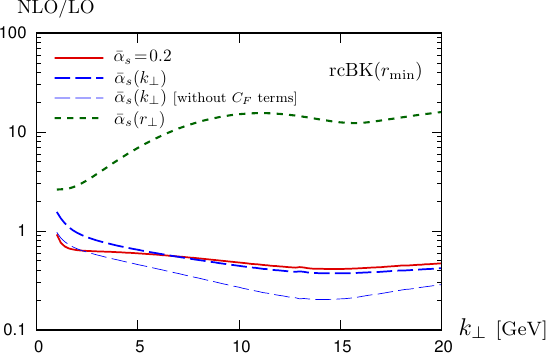}
\caption{Left: Ratio of the NLO multiplicity (including only the $N_c$ terms) and the LO one for different running coupling prescriptions. Right: Ratio of the total NLO quark multiplicity (including both the $N_c$ and $C_F$ terms) and the LO one for three running coupling prescriptions. For comparison we also show the results for $\abar(k_{\perp})$ when including only the $N_c$ terms (same as the curve ``$\abar(k_\perp)$'' in the left panel). For both figures $\sqrt{s}=500~$GeV, $\eta=3.2$ and the evolution of the color dipoles is obtained by solving the BK equation with the smallest dipole prescription. Figure from~\cite{Ducloue:2017dit}.}
\label{fig:daughter}
\end{figure}

Based on this, we arrive at our proposal to use the daughter dipole prescription $\abar(x_{\perp})$ both in the impact factor and when solving the BK equation. Indeed, since $\abar(x_{\perp})$ does not depend on $r_{\perp}$, the final Fourier transform will still eliminate the unphysical contributions~(\ref{jxggr}) coming from the region $x_{\perp} \gg r_{\perp}$. In the physical region $x_{\perp} \sim r_{\perp}$ this choice is equivalent to the parent dipole prescription $\abar(r_{\perp})$ which looks reasonable since $\br$ and $\bk$ are Fourier conjugate. Indeed, the results obtained with this prescription are very close to the ones with a fixed or momentum space coupling as shown in Fig.~\ref{fig:daughter}~(L). Using the daughter dipole prescription throughout the calculation we can thus avoid the ambiguities of the mixed representation. However, note that this prescription is not very natural when solving the BK equation since, as explained previously, one generally expects that the scale of the running coupling should be set by the hardest scale in the problem. Another issue is related to the $C_{\rm F}$ terms and will be discussed in the following section.

\subsection{The $C_{\rm F}$ terms}

For consistency it would be desirable to be able to use the same running coupling prescription throughout the whole calculation. Therefore we now extend the previous discussion to the NLO corrections proportional to the $C_{\rm F}$ color factor. As explained in Sec.~\ref{sec:NLO}, these terms contain collinear divergences which have to be absorbed into the DGLAP evolution of PDFs and fragmentation functions. After this is done, one should replace $\mathcal{I}$~(\ref{I}) and $\mathcal{I}_v$~(\ref{Iv}) by their respective finite expressions $\mathcal{I}^{\rm fin}$~(\ref{Ifin}) and $\mathcal{I}_v^{\rm fin}$~(\ref{Ivfin}). From these expressions one could anticipate that the $C_{\rm F}$ terms are subject to the same fake potential problem as the $N_c$ terms when using a coupling depending on $r_{\perp}$. This is confirmed by the numerical results shown in~\cite{Ducloue:2017dit}. A coordinate running coupling would anyway be inconsistent for these terms since the procedure which was followed in~\cite{Chirilli:2011km,Chirilli:2012jd} to subtract the collinear divergence would not be valid anymore and thus would not lead to the expressions~(\ref{Ifin})-(\ref{Ivfin}). In addition, while for the $N_c$ terms it was possible to alleviate the fake potential issue by using a daughter dipole running coupling $\abar(x_{\perp})$, it is no longer the case here since not all terms in $\mathcal{I}^{\rm fin}$ and $\mathcal{I}_v^{\rm fin}$ can be written as double integrals over $\br$ and $\bx$: the collinear divergence subtraction removes a part of the phase space and we have no control on the daughter dipole size $x_{\perp}$ anymore.

Finally, let us recall that an important feature of the  $C_{\rm F}$ terms is that they vanish in the limit $\xi \to 1$, meaning that the integral over $\xi$ is not logarithmic. This was to be expected since, contrary to the $N_c$ terms, the $C_{\rm F}$ terms are not related to the high energy evolution. While it is easy to see that $\mathcal{I}$ and $\mathcal{I}_v$ vanish separately when $\xi \to 1$, one can show~\cite{Ducloue:2017dit} that the sum of real and virtual $C_{\rm F}$ NLO corrections still vanishes after the collinear divergence subtraction as long as the coupling is fixed or depends on the transverse momentum $k_\perp$. Another important property of the $C_{\rm F}$ (and $N_c$) terms is that they should vanish in the limit $S(\br)=1$, i.e. in the absence of any scattering. As shown in~\cite{Ducloue:2017dit}, these two properties hold with a fixed coupling or a momentum space running coupling $\abar(k_{\perp})$ but would be violated by a coordinate space running coupling such as $\abar(r_{\perp})$.

As a result of these observations, the problem with using a coordinate space running coupling appears to be even more severe in the case of the $C_{\rm F}$ terms than for the $N_c$ terms: the only choice which would look reasonable at first, $\abar(r_{\perp})$, would lead to the fake potential problem discussed in the previous section, generate spurious longitudinal logarithms and prevent the vanishing of the particle production cross section in the absence of scattering. Therefore the only sensible choice for the $C_{\rm F}$ terms seems to be $\abar(k_{\perp})$. This is not fully satisfactory as we cannot use the same coupling in the whole calculation. However, while in the case of the $N_c$ terms it was important to use the same coupling as when solving the BK equation to maintain the equivalence between the subtracted and unsubtracted formulations, there is no such constraint here: the $C_{\rm F}$ terms are a pure NLO correction and the choice of the running coupling scale in these terms should be an NNLO effect. In Fig.~\ref{fig:daughter} we show the NLO quark multiplicity including both the $C_{\rm F}$ and $N_c$ terms for a fixed coupling $\abar=0.2$, $\abar(k_{\perp})$  and the problematic $\abar(r_{\perp})$. To better see the importance of the $C_{\rm F}$ terms we also show, in the case $\abar(k_{\perp})$, the NLO result including only the $N_c$ terms. The effect of including the $C_{\rm F}$ terms is sizable and reduces the difference between the LO and NLO results.

\section{Conclusions}

Promoting the Color Glass Condensate effective theory to NLO accuracy is very important in order to improve the reliability of the predictions in this formalism. Recent progress has been made in this direction by computing the NLO corrections both to the BK evolution and to several process-specific impact factors. The issues met in the first implementations of these corrections have been progressively understood and solved. We focused here on the NLO impact factor for single inclusive forward particle production, which was first found to lead to unphysical results at semi-hard transverse momenta. These issues were due to some approximations which, despite being formally valid at NLO accuracy, can have a large effect because of the necessary cancellation of large contributions between LO and NLO. Physical results can be obtained at fixed coupling by avoiding these approximations, and it should be noted that the calculation of DIS structure functions at NLO is affected by a similar issue, which can be solved in the same way~\cite{Ducloue:2017ftk}.

An issue more specific to the process we considered here is the implementation of the running of the coupling. While the BK equation is usually solved in coordinate space, the impact factor is most naturally expressed in momentum space. This makes the use of a unified running coupling prescription difficult, and using a coordinate space formulation for the impact factor can lead to unphysical results because of the non-commutativity of the Fourier transform with the choice of the running coupling scale. The daughter dipole prescription can overcome most of these issues but it cannot be used for some NLO corrections which are not related to high energy evolution. A possibility to avoid this ambiguity would be to instead solve the BK equation in momentum space, which would allow to use the same transverse scale throughout the whole calculation.

Finally, let us stress that for simplicity we considered here only the quark channel and ignored the fragmentation of partons into hadrons. A realistic calculation should take into account the gluon channel and fragmentation effects. In addition, to reach full NLO accuracy one should also consider the NLO corrections to the high energy evolution of color dipoles~\cite{Balitsky:2008zza,Balitsky:2013fea}, supplemented by the resummation of large collinear logarithms~\cite{Beuf:2014uia,Iancu:2015vea,Iancu:2015joa}.

\section*{Acknowledgments}
This work has been supported by the Agence Nationale de la Recherche, project 
ANR-16-CE31-0019-01.

\end{document}